\documentstyle[11pt,nyas]{article} 

\begin{document}
\title{The Cosmological Mass Function in the Zel'dovich Approximation}
\author{Sergei F. Shandarin}
\affil{ University of Kansas, Lawrence, KS}

\begin{abstract}
The Press-Schechter theory of the cosmological mass function and its
modifications allow to constraint cosmological scenarios of the
structure formation. Recently a few new models have been suggested
that explored the influence of anisotropic collapse on the shape 
of the mass function. I discuss in more detail a particular model
that assumes that a fluid particle becomes a part of a gravitationally
bound halo when the smallest eigenvalue of the deformation tensor of
the filtered initial density field  reaches a certain threshold
(like the filtered density contrast reaches the threshold in the 
Press-Schechter formalism). Choosing the smallest eigenvalue guarantees that
the fluid particle in question experiences collapse along all three
axes. The model shows a  better agreement with the N-body simulations
than the standard Press-Schechter model.
\end{abstract}
\section{Introduction}
The derivation of the distribution of masses in gravitationally bound 
objects is one of the principle goals of the theory of the structure formation
(for a review see \cite{mon98}).
Comparing the theoretical mass function with observations provides 
important constraints on the cosmological models (see e.g. Bond \& Myers 1996,
\cite{bah-fan98}, \cite{rei-etal99}). Rich clusters of galaxies
represent a particular interesting class of objects for two reasons.
Firstly, they are the largest gravitationally bound objects in the universe
and therefore represent rare events. As one of the consequences of being rare
events clusters are particularly sensitive to some parameters of the 
cosmological models ($\Omega_m$ and $\sigma_8$). Secondly, the formation
of clusters is relatively simple process since it is primarily
determined by the gravitational dynamics while other processes (hydro,
thermal, etc) are less important than e.g. in the process of galaxy formation. 
As a result the numerical simulations of clusters of galaxies are more
realistic and reliable than simulations of galaxy formation. 

Measuring the mass function of galaxy clusters is not easy but recently
a certain progress has been achieved for both optically (see e.g. 
\cite{bah-cen93,gir-etal98}) and X-ray (\cite{rei-boe99}) selected samples.
Although there are systematic differences between mass functions obtained
by different groups there is a general agreement in a broad sense.

Most of theoretical derivations of the cosmological mass function are
based on the ideas of \cite{pre-sch74} that can be summarized as follows:
\begin{itemize}
\item The mass fraction ( $F(>M)$ ) in gravitationally bound objects
with masses greater than $M$
can be estimated as the fraction of mass satisfied the collapse condition
at this scale ( $\Psi(\delta_M>\delta_c)$ ): 
$F(>M)= 2 \Psi(\delta_M>\delta_c)$. 
\item The collapse condition is local i.e. it can be expressed in terms of 
the quantities at one point.
\item The quantity that determines the collapse is the linearly 
extrapolated filtered density contrast $\delta_M \ge \delta_c$ at a
given point.
\item The value of the threshold $\delta_c =3/20(12\pi)^{2/3} \approx 1.69$ 
that corresponds
to the collapse of a spherical top-hat model with the similar initial density
contrast. It was assumed that the collapse of the spherical top-hat model
approximately corresponded to the virialization of the gravitationally
bound clump. 
\end{itemize}
The mathematical aspects of the Press-Schechter formalism is outlined in
the following section. Here I would like to discuss briefly some of 
ideas suggested since the formalism was proposed in 1974.

The excursion set approach (\cite{pea-hea90}, \cite{bon-etal91})
justified the assumption that $F(>M)= 2 \Psi(\delta_M>\delta_c)$ in the
case of a sharp $k$-space filter. 

Many realized that the threshold $\delta_c = 1.69$ does not provide the
best fit to N-body simulations. Although some authors used the canonical value
(e.g. \cite{bon-etal91}, \cite{efs-etal88}) others preferred the lower values:
$\delta_c=1.58$ (Bond \& Myers 1996), $\delta_c=1.44$ (\cite{car-cou89}),
or even as low as $\delta_c=1.33$ (\cite{efs-ree88} and \cite{kly-etal95}).
Recently Shapiro et al. (1999) showed that the
virialization of the top-hat model occurs when linear extrapolation
of the density contrast reaches $\delta_c\approx 1.52$.

One of the major efforts in reduction of the discrepancy of the theory 
with simulations has been related to incorporating the anisotropic
character of gravitational collapse. \cite{bon-mye96} developed
a model that incorporated both the Zel'dovich approximation on 
large scales and the collapse of a homogeneous ellipsoid on the
nonlinear scale. \cite{mon95} suggested a different collapse condition
that corresponded to the collapse along the first axis in the Zel'dovich
approximation. \cite{aud-etal97} incorporated some of the nonlinear
effects into an anisotropic collapse model. \cite{lee-sh98a} suggested 
to use the collapse condition 
corresponding to the collapse along all three axes as described by the
extrapolation of the Zel'dovich approximation.
\cite{she-tor99} obtained an analytic fit to the numerical mass 
function in the SCDM, OCDM and $\Lambda$CDM models and then 
\cite{she-etal99} provided a semianalytic derivation of the formula 
assuming an anisotropic collapse an incorporating
some nonlocal effects. All but one models mentioned above
assumed that the formation of a gravitationally bound object is related
to the collapse along three axes. Only \cite{mon95} assumed the collapse
condition corresponding to the collapse along only the first axis. 

In this talk I briefly review the Press-Schechter formalism. Then I
describe the derivation of the mass function ($\lambda_3$-function)
in the Zel'dovich approximation.
I compare the result with the standard Press-Schechter model and
the model suggested by \cite{she-etal99}. I briefly discuss the results
of comparison of the $\lambda_3$-function with N-body simulations.
Finally, I discuss the effect of the initial gravitational potential
on the cosmological mass function and show that the clusters have
a significant tendency to form in the troughs of the initial gravitational
potential. 

\section{The Press-Schechter Formalism}
The mass function $n(M)$ is the number density of gravitationally bound
clumps with masses between $M$ and $M+dM$. 
Let $F(>M)$ be the fraction of the mass contained in the 
gravitationally bound objects with masses greater than $M$. 
Press and Schechter (1974) suggested the fraction $F(>M)$ and 
the mass function $n(M)$ can be related as
\begin{equation}
n(M) = -\frac{\bar{\rho}}{M} \frac{\partial F}{\partial M} \label{mf-F},
\end{equation}
where $\bar{\rho}$ is the mean mass density in the universe and the minus
sign reflects the fact that $F$ is a decreasing function of $M$.

Press and Schechter also made the assumption that
the fraction of mass $F(>M)$ can be estimated as a fraction of
mass $\Psi(\delta_M>\delta_c)$ in the initial density 
field filtered with the window function $W$ (corresponding to $\tilde{W}$
in $k$-space) 
\begin{equation}
\delta_M({\bf x},t)=D(t) \int \delta_{in}({\bf x'})
W(|{\bf x'}-{\bf x}|/R)~d^3x'
\end{equation}
satisfying the collapse condition $\delta_M > \delta_c$. Here 
$\delta=(\rho-\bar{\rho})/\bar{\rho}$ is the density contrast, 
$D(t)$ is linear growth factor, ${\bf x}$ is the comoving coordinate. 
The mass $M$ and the linear scale $R$ of the filter are related as
\begin{equation}
M=f_W \frac{4\pi}{3}R^3\bar{\rho},
\end{equation}
where $f_W$ is a factor depending on the shape  of the 
smoothing filter $W$. For a sharp k-space filter adopted here
$f_W=9\pi/2 \approx 14.1$ and thus $M=6\pi^2 R^3\bar{\rho}$
(see e.g. \cite{lac-col94}).  

Assuming that the initial density contrast is Gaussian random field
its pdf (probability distribution function) is
\begin{equation}
p(\delta_M) = \frac{1}{\sqrt{2\pi}\sigma_M}
\exp\bigg{[}-\frac{\delta_M^2}{2\sigma^2_M}\bigg{]}, \label{p_del}
\end{equation}
where the variance $\sigma^{2}_M$ is a function of mass $M$
\begin{equation}
\sigma^{2}_M = \int\! \frac{d^3k}{(2\pi)^3}
P(k) \tilde{W}^{2}(kR) , \label{variance}
\end{equation}
where $P(k) =|\delta_{k}|^2$ is the initial spectrum of perturbations
and $\tilde{W}(kR)$ is the window function in the k-space.

Press and Schechter argued that a fluid element becomes a part of
a gravitationally bound object of mass $M$ when its linearly
extrapolated density contrast $\delta _M$
reaches the critical value $\delta_{c}=3/20(12\pi)^{2/3} \simeq 1.69$. 
This corresponds to the collapse of the top-hat spherical perturbation
having the initial density contrast similar to the fluid element in question.
The collapse of a spherical top-hat model has been assumed approximately
to correspond the virialization of the clump. Recently Shapiro et al. (1999)
showed that the virialization corresponds to  $\delta_{c} \simeq 1.52$ rather
than to  $\delta_{c} \simeq 1.69$.

The fraction of mass satisfying the collapse condition on scale $M$ is
\begin{eqnarray}
\Psi(\delta_M>\delta_c)
&=& \frac{1}{\sqrt{2\pi}\sigma(M)}\int^{\infty}_{\delta_{c}}\!
\exp\bigg{[}-\frac{\delta_M^2}{2\sigma^2(M)}\bigg{]}d\delta_M \nonumber \\
&=& \frac{1}{2}{\rm erfc}\bigg{[}\frac{\delta_{c}}{\sqrt{2}\sigma(M)}
\bigg{]}, \label{mass_fr}
\end{eqnarray}
where erfc$(x)$ is the complementary error function. Assuming that 
$F(M) \approx \Psi(\delta_M>\delta_c)$ one easily obtains the mass function 
$n(M)$ (eq. \ref{mf-F} ).

One obvious problem with this result is that the normalization integral
\begin{equation}
\int^{\infty}_{0}dF \approx \Psi(\delta_{M=\infty}>\delta_c) = \frac{1}{2}
\end{equation}
meaning that only a half of the mass is contained in the gravitationally 
bound clumps.
Press and Schechter renormalized $n(M)$ by introducing an additional
factor of 2 ($F(M) = 2 \Psi(\delta_M>\delta_c$ ) 
\begin{eqnarray}
n_{ps}(M) &=& -2\frac{\bar{\rho}}{M}
\frac{\partial \Psi}{\partial M} 
 = 2\frac{\bar{\rho}}{M}\frac{d\sigma}{dM}
\frac{\partial \Psi}{\partial \sigma} \nonumber \\
&=& -\sqrt{\frac{2}{\pi}}
\frac{\bar{\rho}}{M}\frac{d\sigma}{dM}
\frac{\delta_{c}}{\sigma^2(M)}\exp\bigg{[}-\frac{\delta_{c}^2}
{2\sigma^2_M}\bigg{]}.       \label{ps_function}
\end{eqnarray}

Later, the normalization problem was correctly resolved in the frame 
of the excursion set model (\cite{pea-hea90} and 
\cite{bon-etal91}). 
The derivation of Press and Schechter did not take into account
the so called cloud-in-cloud problem.
Function $\Psi(\delta_M>\delta_c)$ in eq.\ref{mass_fr} gives 
the fraction of mass
that satisfies the collapse condition at the filtering scale $M$. However,
some of the fluid particles may satisfy the collapse condition
at larger filtering scales. In the correct model the fluid
elements must be assigned to the clumps of mass $M_1$ being equal to
the largest filtering mass at which the collapse condition 
is fulfilled. In the
excursion set formalism  this corresponds to the first crossing of
the collapse threshold $\delta_c$ while $\delta$ evolves with the 
growth of $\sigma$.

An elegant method to normalize the mass function was suggested 
by \cite{jed95} (see also the discussion in \cite{yan-etal96}) 
who derived the integral equation
\begin{equation}
\Psi(\delta_M>\delta_c) = \int^{\infty}_{M}\!dM_1
n(M_1)\frac{M_1}{\bar\rho}P(M,M_1). \label{int_eq}
\end{equation}
that relates the fraction of the fluid elements firstly crossed the collapse
threshold at the filtering scale $M_1$ ( $dM_1n(M_1)M_1/{\bar\rho}$ ) and
the fraction of mass satisfying the collapse condition at filtering
scale $M$ ( $\Psi(\delta_M>\delta_c)$ ). 
Function $P(M,M_1)$ is the probability 
that a fluid particle firstly crossed the collapse threshold at the 
scale $M_1$ satisfies
the collapse condition at the scale $M$. In the case of the
sharp k-space filter and Gaussian $\delta_M$ this probability
is exactly equal to $1/2$ for all $M_1>M$. Thus, the integral equation
(\ref{int_eq}) can be immediately solved for the mass function $n(M)$.
The solution is the correctly normalized mass function of 
eq. \ref{ps_function}.

\section{Mass Function in the Zel'dovich Approximation}
The simplest theory describing the anisotropic character of the 
gravitational collapse
in a generic case of random initial condition is the Zel'dovich
approximation (\cite{zel70}, see also \cite{sh-zel89} for a
discussion). In particular, the Zel'dovich approximation provides 
a formula for an anisotropic collapse of a fluid element
\begin{equation}
\rho({\bf q},t) = \frac{\bar{\rho}}
{[1 - D(t)\lambda_{1}({\bf q})][1 - D(t)
\lambda_{2}({\bf q})][1 - D(t)\lambda_{3}({\bf q})]} , \label{z_appr}
\end{equation}
where $D(t)$ is the linear growth function and $\lambda_{1}({\bf q}), \lambda_{2}({\bf q})$ and $\lambda_{3}({\bf q})$ are the eigenvalues of the initial 
deformation tensor. Using the ordering convention $\lambda_{1}({\bf q}) > \lambda_{2}({\bf q})$ and $\lambda_{2}({\bf q}) > \lambda_{3}({\bf q})$  
the condition $1 - D(t)\lambda_{1}({\bf q})=0$ has been interpreted 
as a collapse of a fluid particle along one principle axis (\cite{zel70}).
Similarly the conditions $1 - D(t)\lambda_{i}({\bf q})=0$ ($i=2,3$) can
be interpreted as collapses along the second and third principle axes.

\cite{sh-kly84} showed that the formation of gravitationally bound clumps
was the best correlated with the maxima of the smallest 
eigenvalue ($\lambda_3$) of the initial deformation tensor. Although
the formation of the clumps may be also related to other pointlike
singularities (\cite{arn-etal82}) here we assume that a fluid particle
becomes a part of a gravitationally bound clump of mass $M$ when 
its smallest eigenvalue $\lambda_3$ reaches the critical value
$\lambda_c$ at the largest filtering scale $M$ (Lee \& Shandarin 1998a).
The Zel'dovich approximation (eq.\ref{z_appr}) predicts that the 
collapse condition is $\lambda_c=1$ (it is assumed that $D(t)$ normalized to
$D(t_0)=1$, where $t_0$ is the present time). 
However, because of multistreaming
effect all fluid particles (except the set of measure zero) enter the
multi-stream flow regions before they collapse. We approximately incorporate
this complex effect by reducing the threshold $\lambda_c$ to a smaller value.
The comparison with the Press-Schechter mass function as well as with
the numerical mass function suggests that $\lambda_c=0.37$ is a reasonable
choice. 

The derivation of the mass function in the Zel'dovich approximation 
is similar to the Press-Schechter
derivation except that the collapse condition is $\lambda_3(M)=\lambda_c$
instead of $\delta_M=\delta_c$.
\cite{dor70} derived the joint pdf of three eigenvalues
\begin{equation}
p(\lambda_{1},\lambda_{2},\lambda_{3}) = 
\frac{3375}{8\sqrt{5}\pi\sigma^6}\exp\bigg{(}-\frac{3I_{1}}{\sigma^2}
 + \frac{15I_{2}}{2\sigma^2}\bigg{)}(\lambda_{1}-\lambda_{2})
(\lambda_{2}-\lambda_{3})(\lambda_{1}-\lambda_{3}) , \label{pdf_3lam}
\end{equation}
where $I_{1} = \lambda_{1}+\lambda_{2}+\lambda_{3}$, $I_{2} = 
\lambda_{1}\lambda_{2} + \lambda_{2}\lambda_{3} + \lambda_{3}\lambda_{1}$ 
and $\sigma^2$ is the density contrast variance as defined 
in eq. \ref{variance}.
Integrating $p(\lambda_{1},\lambda_{2},\lambda_{3})$ 
over two eigenvalues one can obtain the pdf
of one of the eigenvalues. We are interested in the collapse along the third 
axis and therefore $p(\lambda_{3})$ is of primary interest  
\begin{eqnarray}
p(\lambda_{3}) & = &\frac{\sqrt{5}}{12\pi\sigma} \Bigg{\{}
3\sqrt{3\pi}\exp\bigg{(}-\frac{15\lambda_{3}^2}{4\sigma^2}\bigg{)}
{\rm erfc}\bigg{(}\frac{\sqrt{3}\lambda_{3}}{2\sigma}\bigg{)}\nonumber \\
&+& \sqrt{2\pi}\bigg{(}20 \frac{\lambda_{3}^2}{\sigma^2}-1\bigg{)}
\exp\bigg{(}-\frac{5\lambda_{3}^2}{2\sigma^2}\bigg{)}
{\rm erfc}\bigg{(}\sqrt{2}\frac{\lambda_{3}}{\sigma}\bigg{)} \nonumber \\
&-& 20\frac{\lambda_3}{\sigma}
\exp\bigg{(}-\frac{9\lambda_{3}^2}{2\sigma^2}\bigg{)}\Bigg{\}}.\label{p_lam3}
\end{eqnarray}
Repeating the derivation of the previous section using the pdf of
 eq.\ref{p_lam3} instead of
eq.\ref{p_del} one arrives to an analog of the normalization integral 
equation (eq.\ref{int_eq})
\begin{equation}
\Psi(\lambda_3(M)>\lambda_c) = \int^{\infty}_{M}\!dM_1
n(M_1)\frac{M_1}{\bar\rho}P(M,M_1), \label{int_eq_l}
\end{equation}
here $\Psi(\lambda_3(M)>\lambda_c)$ is the fraction of mass where 
$\lambda_3(M)>\lambda_c$ on the filter scale $M$.

Solving exactly eq.\ref{int_eq_l} is much more difficult than eq.\ref{int_eq} 
because now $P(M,M_1)$ is not a constant. 
In the limit $M_1-M \ll M$ the probability $P=0.5$ as in eq.\ref{int_eq} 
but in the limit $M_1 \gg M$ it drops to $P=0.08$. 
\cite{lee-sh98a} used the limiting value $P=0.08$ and analytically derived 
the  mass function in the Zel'dovich approximation
\begin{eqnarray}
n(M) &=&-\frac{25\sqrt{5}}{24\pi}
\frac{\bar{\rho}}{M}\frac{d\sigma}{dM}
\frac{\lambda_{3c}}{\sigma^2_M}\nonumber \\
&\Bigg{\{}&
3\sqrt{3\pi}\exp\bigg{(}-\frac{15\lambda_{3}^2}{4\sigma^2}\bigg{)}
{\rm erfc}\bigg{(}\frac{\sqrt{3}\lambda_{3}}{2\sigma}\bigg{)}\nonumber \\
&+& \sqrt{2\pi}\bigg{(}20 \frac{\lambda_{3}^2}{\sigma^2}-1\bigg{)}
\exp\bigg{(}-\frac{5\lambda_{3}^2}{2\sigma^2}\bigg{)}
{\rm erfc}\bigg{(}\sqrt{2}\frac{\lambda_{3}}{\sigma}\bigg{)}\nonumber \\
&-& 20\frac{\lambda_3}{\sigma}
\exp\bigg{(}-\frac{9\lambda_{3}^2}{2\sigma^2}\bigg{)} \Bigg{\}}. 
\end{eqnarray}

In the following sections I compare the obtained result with the
Press-Schechter and Sheth-Mo-Tormen mass functions as well as with
numerical simulations.

\section{Comparison of Three Analytic Mass Functions}
Both the Press-Schechter and $\lambda_3$-mass functions have a common factor
$\frac{\bar{\rho}}{M}\frac{d\sigma}{dM}$ which depends on the initial spectrum
and $f(\sigma) \equiv \partial F/\partial \sigma$ 
that completely characterizes a model. 
Thus, comparing different models is convenient by comparing 
$f(\sigma)$ as functions of $\sigma$. The Press-Schechter and 
$\lambda_3$-functions are
\begin{eqnarray}
f_{PS}(\sigma) 
&=&\sqrt{\frac{2}{\pi}}
\frac{\delta_{c}}{\sigma^2}\exp\left(-\frac{\delta_{c}^2}
{2\sigma^2}\right), \nonumber \\
f_{\lambda_3}(\sigma) 
&=& \frac{25\sqrt{5}}{24\pi}\frac{\lambda_{3c}}{\sigma^2}\Bigg{\{}
3\sqrt{3\pi}\exp\bigg{(}-\frac{15\lambda_{3}^2}{4\sigma^2}\bigg{)}
{\rm erfc}\bigg{(}\frac{\sqrt{3}\lambda_{3}}{2\sigma}\bigg{)}\nonumber \\
&+& \sqrt{2\pi}\bigg{(}20 \frac{\lambda_{3}^2}{\sigma^2}-1\bigg{)}
\exp\bigg{(}-\frac{5\lambda_{3}^2}{2\sigma^2}\bigg{)}
{\rm erfc}\bigg{(}\sqrt{2}\frac{\lambda_{3}}{\sigma}\bigg{)}\nonumber \\
&-& 20\frac{\lambda_3}{\sigma}
\exp\bigg{(}-\frac{9\lambda_{3}^2}{2\sigma^2}\bigg{)} \Bigg{\}}.
\end{eqnarray} 
Sheth and Tormen (1999) and Sheth, Mo and Tormen (1999) derived a new 
mass function that fits better the results of the N-body simulations
\begin{equation}
f_{SMT}(\sigma) 
= A\left[ 1+\left(\frac{a\delta_c^2}{\sigma^2}\right)^{-q} \right]
\sqrt{\frac{2}{\pi}}\frac{\sqrt{a}\delta_c}{\sigma^2}
\exp\left(-\frac{a\delta_c^2}{\sigma^2}\right),
\end{equation}
here $a=0.707$, $q=0.3$ and the constant $A=0.322$ found from the 
normalization condition 
\begin{equation}
\int_0^{\infty}f(\sigma)d\sigma=1. \label{norm}
\end{equation}
Choosing 
$a=1$, $q=0$ and the constant $A=1$ one obtains the Press-Schechter
function. These three functions are shown in Fig. 1a. The small box
shows the range of $\sigma$ where the theoretical mass functions were
checked against N-body simulations by \cite{she-tor99} and \cite{lee-sh99}.
Fig. 1b shows the ratios $f_{PS}/f_{SMT}$  and $f_{\lambda_3}/f_{SMT}$.

\section{Comparison with  N-body Simulations}
Figure 2 shows the comparison of the $\lambda_3$-function with the
numerical mass functions for the scale invariant initial spectra:
$P(k) \propto k^n$ with $n=-1$ and $n=0$ (see for the details
\cite{tor98}). The top panel
($n=-1$) shows a quite good agreement of the $\lambda_3$-function with  
the numerical mass function, while in the $n=0$ case the agreement is 
significantly worse. Fig. 3 shows the comparison of the $\lambda_3$-function
with the N-body simulation of the SCDM model (\cite{gov-etal99}).
At four epochs ($z=1.86, 1.14, 0.43$, and $0$) the $\lambda_3$-function
is in a better agreement that the Press-Schechter mass function.
A similar result has been reported by \cite{she-tor99}.

\section{Large-Scale Biasing}
It has been noticed for sometime that the initial gravitational 
potential may noticeably affect the large scale structure.
Kofman and Shandarin (1988) showed that the adhesion approximation 
predicts that the formation of voids 
is associated with positive peaks of the primordial 
gravitational potential. Sahni et al. (1994)
studied the effect  and measured a significant correlation between 
the sizes of voids and  the value of primordial gravitational potential
in numerical simulations of the adhesion model. 
Recently, Madsen et al. (1998) have demonstrated by N-body  
simulations that the underdense and the overdense regions are 
closely linked to the regions with the positive and the negative 
gravitational potential respectively. \cite{lee-sh98b} showed that
the initial potential also affects the masses of clusters. 

In order to incorporate the 
primordial gravitational potential fluctuations
term into the derivation of the mass function, 
we first derive the conditional probability density distribution 
$p(\delta|\varphi<-\varphi_{c})$ ($\varphi_{c}>0$):
\begin{eqnarray}
p(\delta|\varphi<-\varphi_{c}) &=& 
\frac{1}{\sqrt{2\pi}\sigma_{\delta}}
\exp\Bigg{(}-\frac{\delta^2}{2\sigma_{\delta}^2}\Bigg{)}
\Bigg{[}1-{\rm erf}\bigg{(}
\frac{\varphi_{c}}{\sqrt{2}\sigma_{\varphi}}
\bigg{)}\Bigg{]}^{-1} \times \nonumber \\ 
&&\Bigg{[}1+{\rm erf}\Bigg{(}\frac{\kappa
\frac{\delta}{\sigma_{\delta}}- \frac{\varphi_c}{\sigma_{\varphi}}}
{\sqrt{2(1-\kappa^2)}}
\Bigg{)}\Bigg{]}.
\end{eqnarray}
Here $\sigma_{\delta}^2$, $\sigma_{v}^2$, and $\sigma_{\varphi}^2$
are the density, velocity and the potential 
variances respectively; $\kappa = <\delta\cdot\varphi>/ 
\sigma_{\delta}\sigma_{\varphi} =
\sigma_{v}^2/\sigma_{\delta}\sigma_{\varphi}$ is 
the crosscorrelation coefficient of the the density contrast 
$\delta$ smoothed on the scale $k_c$ and the primordial 
({\it unsmoothed}) potential fluctuations $\varphi$.
As a result, eq.\ref{mf-F} for the conditional mass function 
$n(M|\varphi<-\varphi_{c})$ becomes
\begin{equation}
n(M|\varphi<-\varphi_{c}) = -\frac{\bar{\rho}}{M}
\left(\frac{\partial F}{\partial\sigma_{\delta}}
\frac{d\sigma_{\delta}}{dM} 
+ \frac{\partial F}{\partial\sigma_{v}}
\frac{d\sigma_{v}}{dM}\right).
\end{equation}
The further calculation needs to be done numerically. Fig. 4 
illustrates how the mass function depends on the initial potential
in  the CDM model
with $\Gamma = \Omega h = 0.25$ normalized to $\sigma_{8} = 1$.
The top panel shows the mass function for regions of positive and
negative initial potential as well as unconditional mass function.
The bottom panel show the ratio of conditional mass functions
to unconditional one.

We also calculate the probability that a clump
with mass M is located in the potential regions satisfying the 
chosen condition, for instance,  $\varphi<-\varphi_{c}$  
\begin{equation}
P(\varphi<-\varphi_{c}|M) = \frac{n(M|\varphi<-\varphi_{c})}
{n(M)}P(\varphi<-\varphi_{c}),
\end{equation}
where $P(\varphi<-\varphi_{c})$ is the fraction of space
satisfying the given condition (see Fig. 5).

The scale of the initial potential 
\begin{equation}
R_{\varphi}
= \sqrt{3} \sigma_{\varphi}/
\sigma_{\varphi'} 
=\sqrt{3\frac{\int^{\infty}_{k_{l}}\! dk k^{-2}P(k)}
{\int^{\infty}_{0}\! dk P(k)}}
\approx 120 h^{-1} {\rm Mpc}
\end{equation}
does not depend on any ad hoc scale; the dependence on $k_l$ is
extremely weak ($\propto \sqrt{ln{(1/k_l)}}$ for the Harrison-Zel'dovich
spectra assumed here).
The geometry of the gravitational potential does not evolve much on 
large scales (\cite{kof-sh88}, \cite{mad-etal98}).  
Therefore, the potential
at present is very similar to the primordial one on scales much greater
than the scale of nonlinearity. 
A simple explanation to this in the frame of the standard scenario
of the structure formation is due to the fact that the mass 
has been displaced by the distance about $10 h^{-1}{\rm Mpc}$
(\cite{sh93}). Therefore, the potential on  scales
greater than, say, $30 h^{-1}{\rm Mpc}$ has been changed very little.

For the model in question the scale of the primordial potential
is found to be $R_{\varphi} \approx 120 h^{-1} {\rm Mpc}$. 
The scale of the density contrast field reaches this value 
$R_{\delta} = \sqrt{3} \sigma_{\delta}/\sigma_{\delta'} 
\approx 120 h^{-1} {\rm Mpc}$ 
only after it is smoothed  on $k_c \approx 0.017 h {\rm Mpc^{-1}}$.  
The corresponding density variance on this scale is 
$\sigma_{\delta}(0.017 h {\rm Mpc^{-1}}) \approx 0.03$.
On the other hand, the number of clumps
with masses $10^{14} - 10^{15} h^{-1} M_{\odot}$ can
easily be  30\% 
greater in the troughs of the potential than the mean density 
$n(>M) = 0.5[n(>M|\varphi<0)+ n(>M|\varphi>0)]$ 
(see Fig. 4). Thus, the bias factor $b$ (defined by the relation
$\Delta n_{cl}/ n_{cl} = b \Delta \rho_m/\rho_m$) 
reaches at least $10$ on the scale about $120 h^{-1} {\rm Mpc}$.

Figure 5 demonstrates that the most massive clusters 
($M>10^{15}h^{-1}M_{\odot}$) are almost certainly located in the
the troughs in the initial potential. The bias defined as the density
contrast of the clusters with respect to the mass density contrast
$b=\delta_{cl}/\delta{\rho}$ reaches the value $3-10$ on the scale 
of the potential $R_{\phi} \approx 120h^{-1}Mpc$ (Lee \& Shandarin 1998b).

\section{Summary}
In the talk I discussed new modifications of the Press-Schechter 
theory of the cosmological mass function. One assumes a different
collapse condition that implies that a fluid particle becomes a part
of a gravitationally bound object after it experiences collapses
along three axes. The comparison with other models (Fig. 1) shows
that it predicts about 25\% more gravitationally bound clumps than
the Sheth-Mo-Tormen model in the range $ 0.45 \ge \sigma \ge 3.1$
where the comparison with the N-body simulations has been done.
A direct comparison with the N-body simulations (Fig. 2 and 3) 
shows a quite good agreement although not as good as the 
Sheth-Mo-Tormen model. The $\lambda_3$-function based on the
Zel'dovich approximation has been obtained analytically similarly
to the Press-Schechter function. A drawback of the derivation is
a quite crude approximation of the probability function $P(M,M_1)$
in the normalization integral eq.\ref{int_eq_l}. A more
accurate normalization will be reported separately. The Sheth-Mo-Tormen 
model also suffer from a normalization problem: the shape of the
mass function has been derived but the normalization has been enforced
by demanding equality of eq.\ref{norm} 

Another modification is the conditional mass function showing that
the clusters of galaxies tend to form in the troughs of the initial
gravitational potential and avoid the peaks of the potential. 
The gravitational potential field has a typical scale of 
about $120h^{-1}Mpc$ and as a result has an
advantage of being independent of the arbitrariness of the smoothing scale
(if the filter scale is smaller than roughly $50h^{-1}Mpc$) and
at present it has almost same geometry as at the epoch of decoupling.
Figures  4 and 5 quantify this large-scale biasing.\\

\section{Acknowledgments} 
I am grateful to Ravi Sheth for useful discussions
during the workshop. This work has been partly supported by the University of
Kansas GRF 2001 grant.

\vfill
\newpage
\begin{center}{\bf Figure Captions}\end{center}

Fig. 1. (a) The fraction of mass $f=dF/d\sigma$ in the gravitationally 
bound objects as a function of $\sigma$ as predicted by 
the Press-Schechter model
(short dashed line), Sheth-Mo-Tormen model (solid line), and $\lambda_3$-model
(long dashed line). The small box shows the range of $\sigma$ where
the models have been checked against N-body simulations (see Fig. 2 in
\cite{she-etal99})\\ 
(b) The logarithm of the ratios $f_{PS}/f_{SMT}$ (short-dashed line) and 
$f_{\lambda_3}/f_{SMT}$ (long-dashed line).\\ 

Fig. 2. The square dots represent the numerical mass
function with poissonian error bars. The solid line is the $\lambda_3$-mass 
function with $\lambda_{3c} = 0.37$ while 
 the dashed, the dotted lines are the PS mass functions with 
$\delta_{c} = 1.69, 1.5$ respectively.  The upper and the lower 
panels correspond to the $n=-1$ and the $n=0$ power-law models respectively. 
See also the top left panel of Fig.2 in \cite{tor98}. \\

Fig. 3. The square dots represent the numerical data 
for the case of SCDM model with $\Omega = 1, h=0.5$. 
The solid line is our mass function with $\lambda_{3c} = 0.37$, 
and the dashed, the dotted lines
are the PS mass functions with $\delta_{c} = 1.69, 1.5$ respectively.\\

Fig. 4. In the upper panel the conditional cumulative  
mass function satisfying chosen potential condition is plotted. 
The solid, the long dashed, the 
dot-dashed, and the dashed lines correspond to the conditions  
$\varphi  < -\sigma_{\varphi}$, $\varphi  < 0$, 
$\varphi  > 0$, and  $\varphi  > \sigma_{\varphi}$ respectively, 
while the dotted line represents the unconditional cumulative 
PS mass function.  The shaded area is $1\sigma$ fit to 
the observational cumulative mass function of rich clusters
by Bahcall and Cen (1993).
In the lower panel the ratio of the 
conditional  cumulative mass functions to the unconditional 
one is plotted for each condition. 
The CDM spectrum with $\Gamma = 0.25$ normalized to 
$\sigma_{8} = 1$ has been used. \\

Fig. 5. The probability that a
clump with mass M can be found in the regions satisfying chosen
potential condition is plotted. The heavy solid, the 
heavy dashed, the solid, the dashed, the 
long dashed, and the dot-dashed lines correspond to the 
condition $\varphi  < 0$, $\varphi  > 0$,
$\varphi  < -\sigma_{\varphi}$, $-\sigma_{\varphi} < \varphi  < 0$, 
$0 < \varphi < \sigma_{\varphi}$, and   
$\varphi  > \sigma_{\varphi}$ respectively.
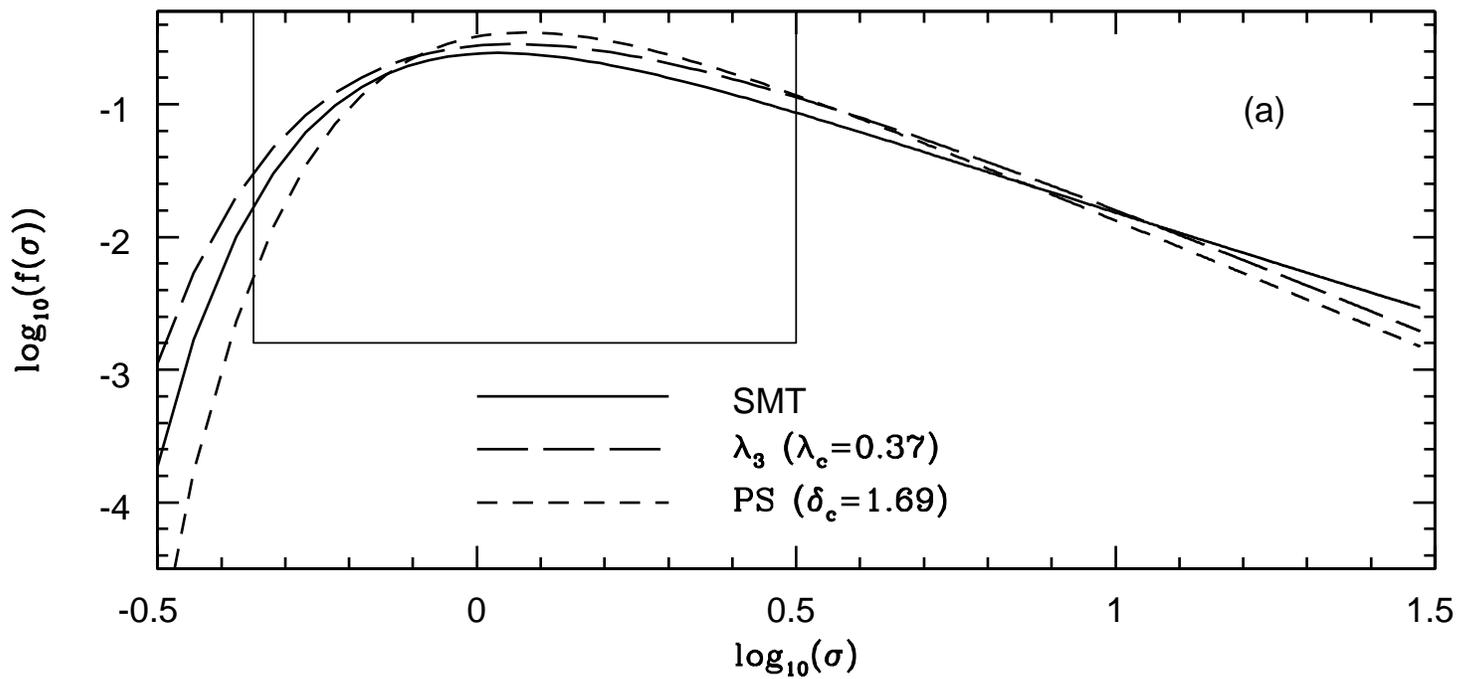
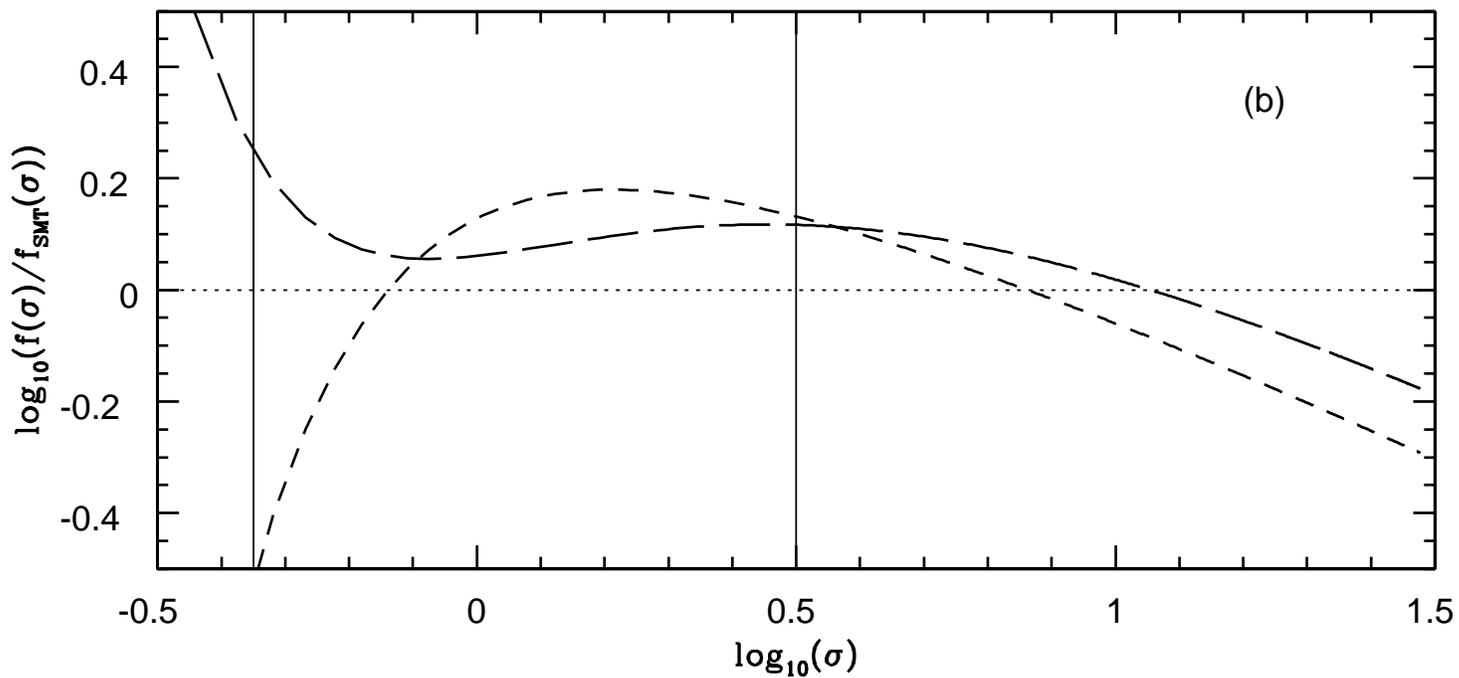
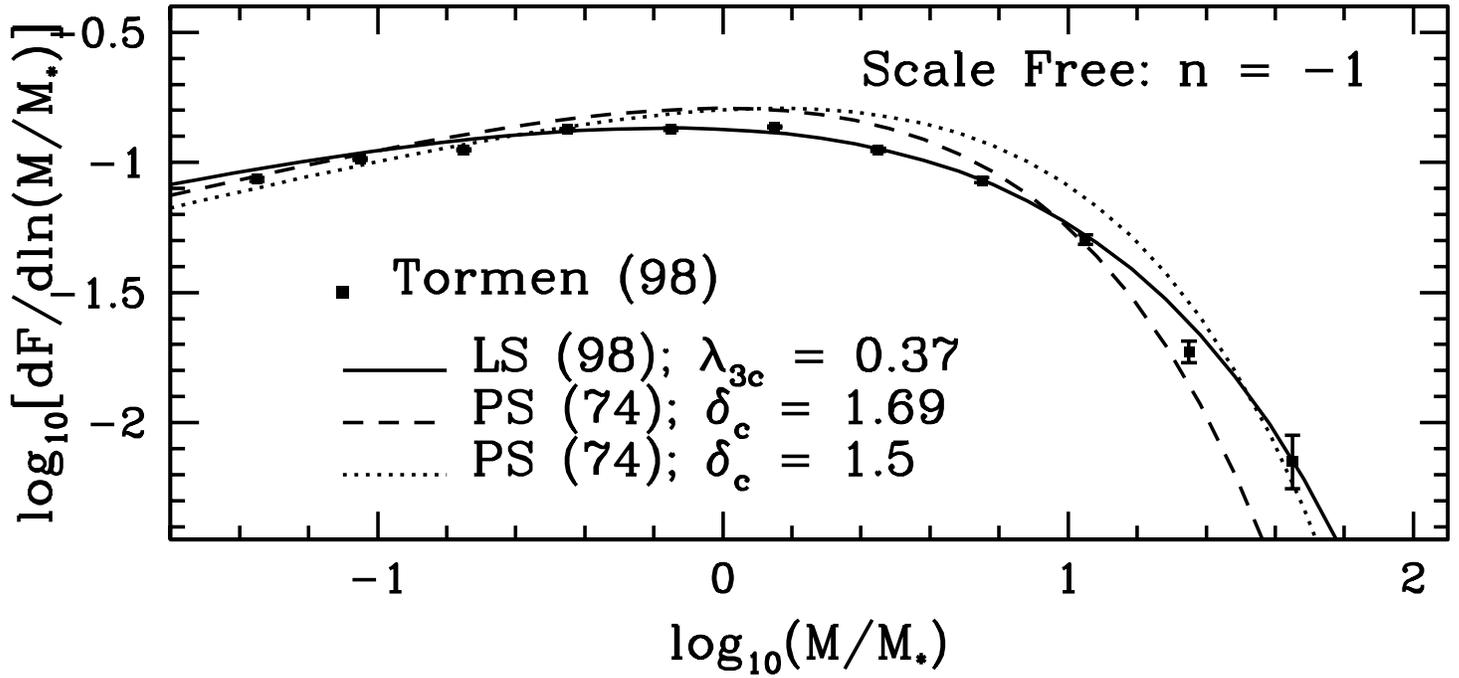
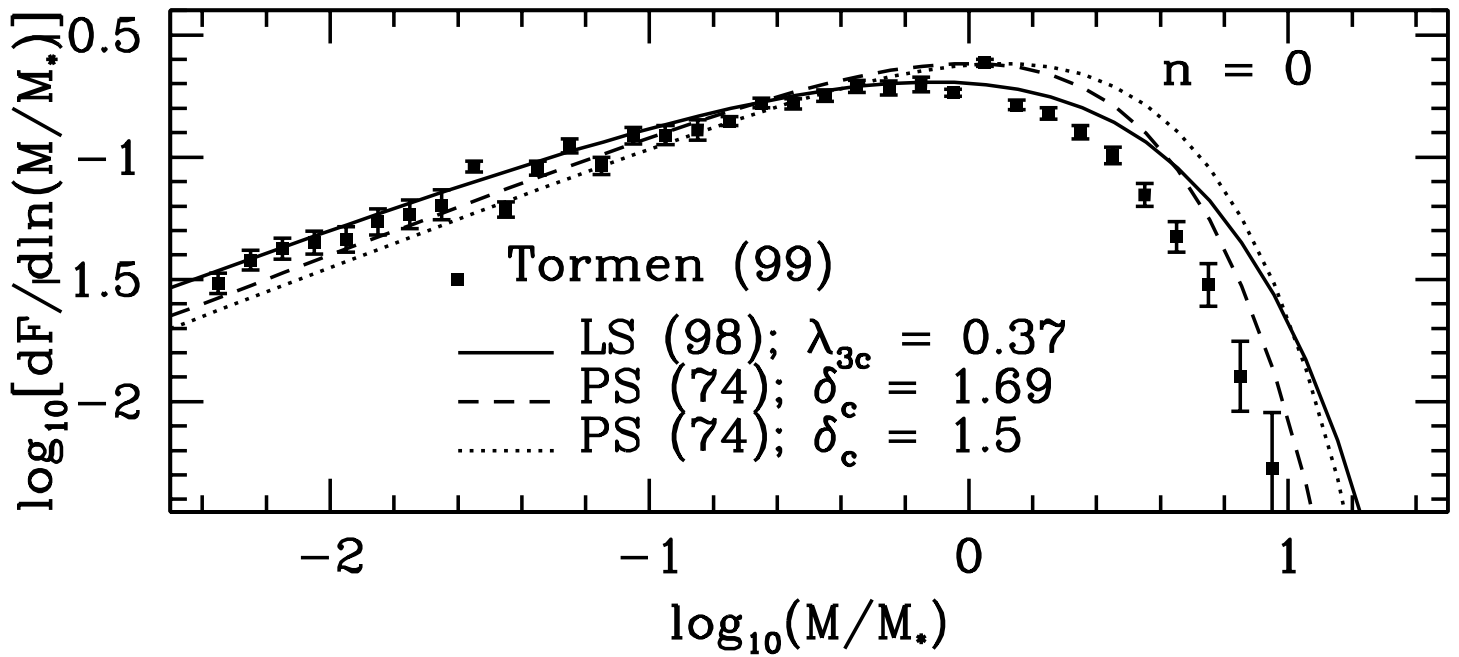
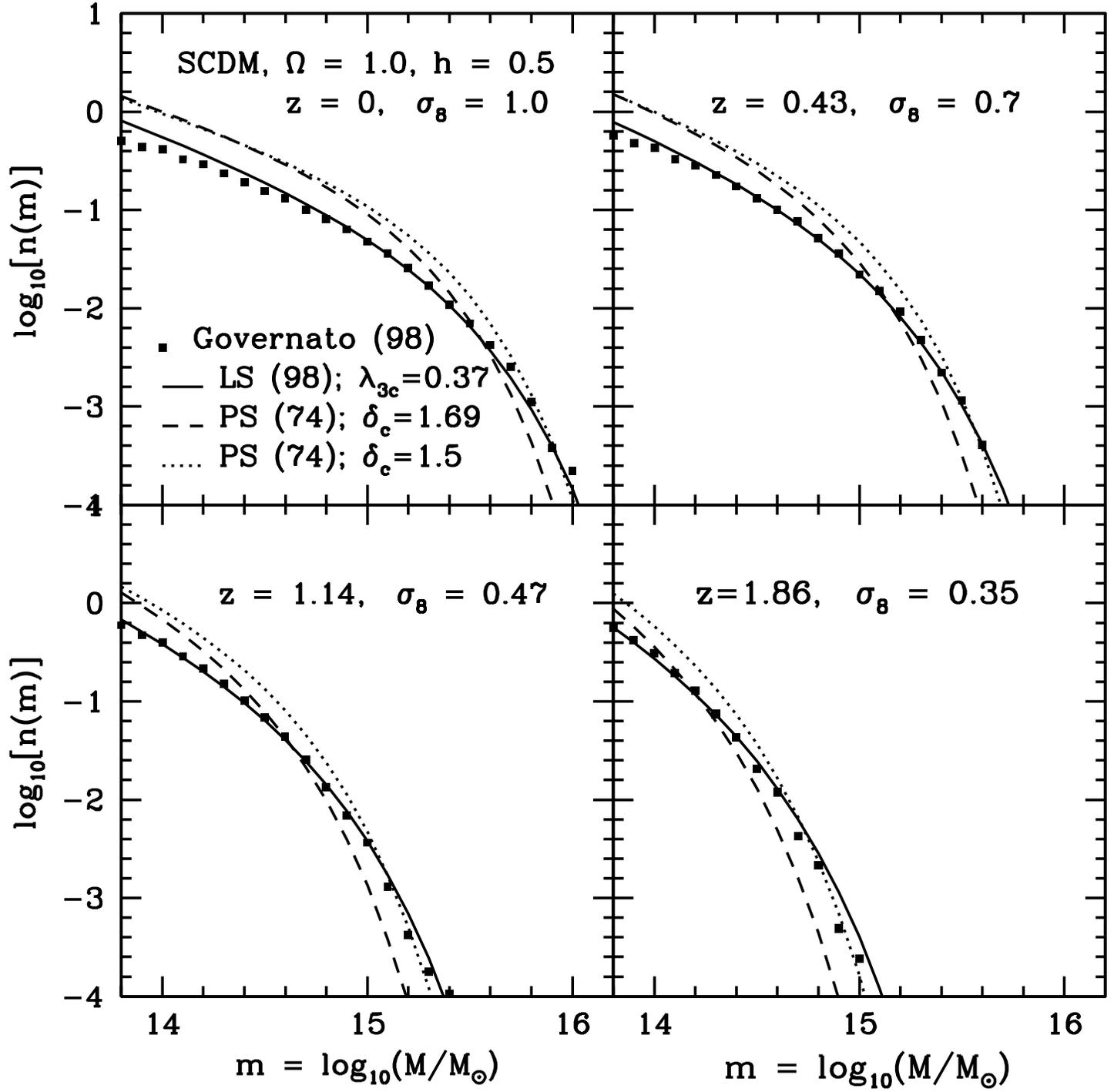
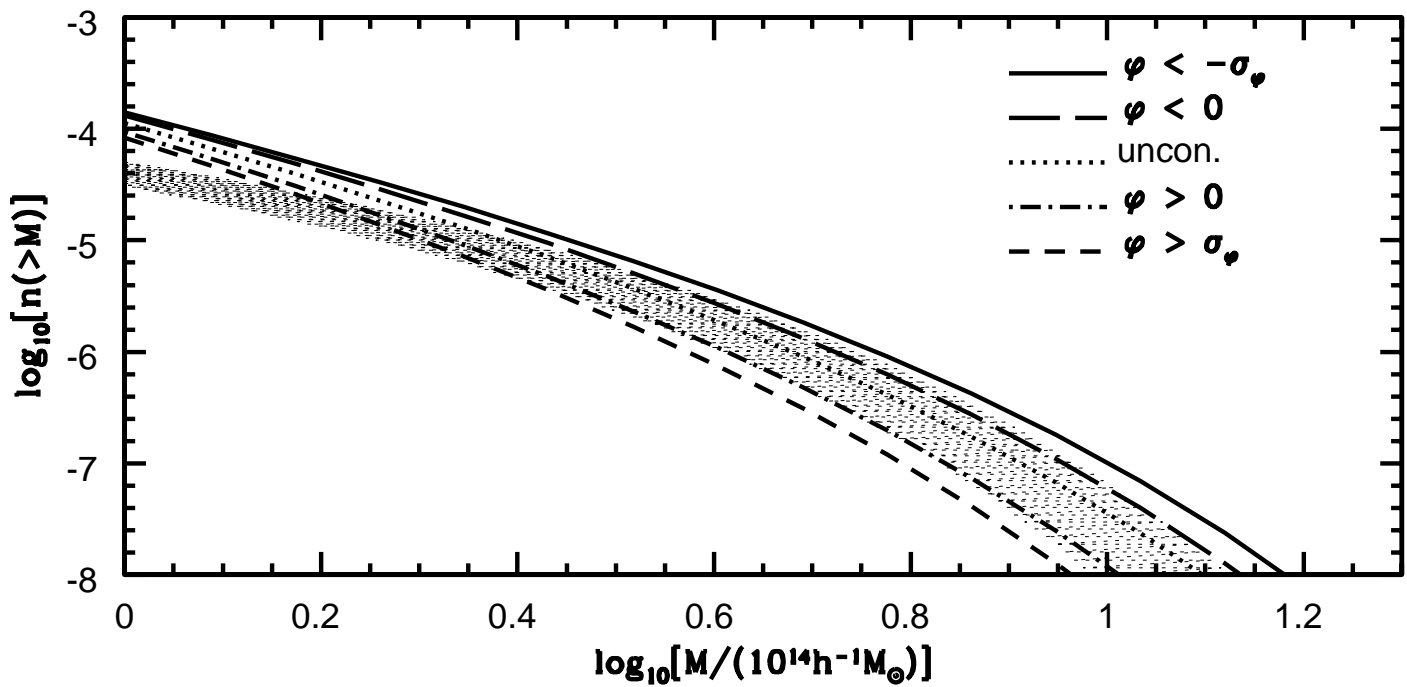
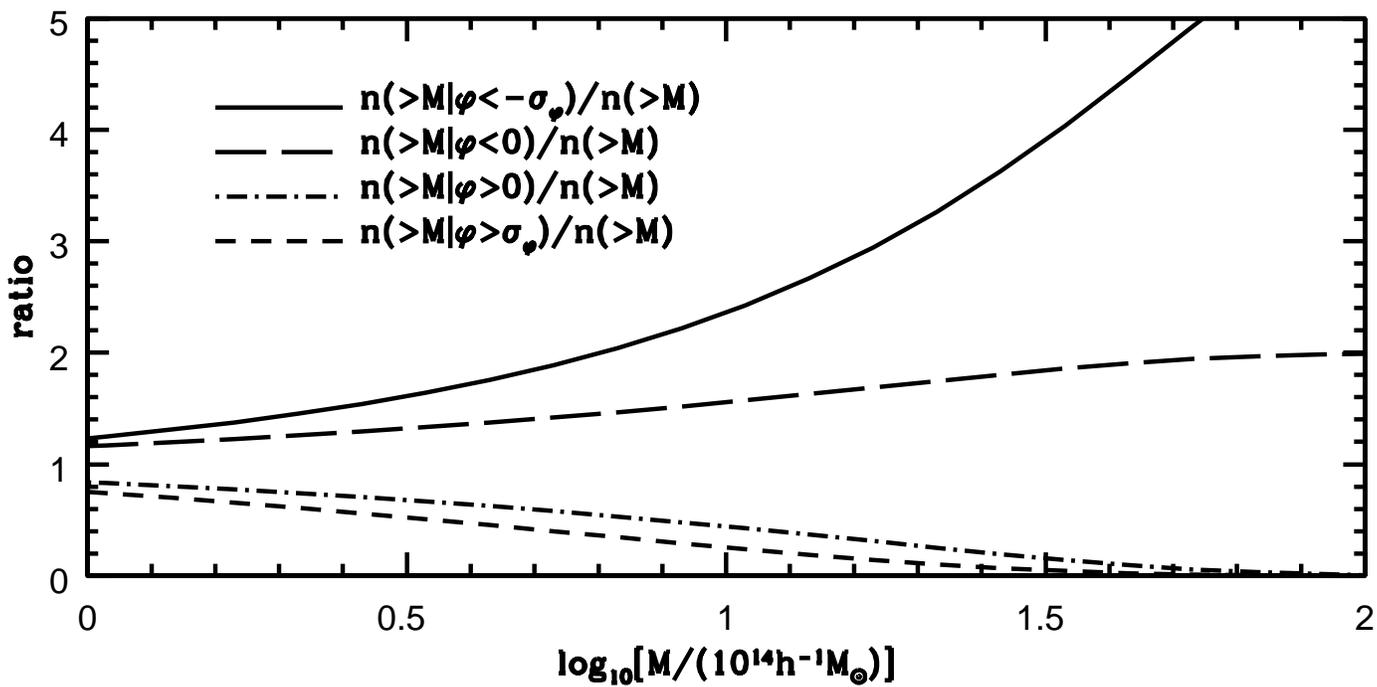
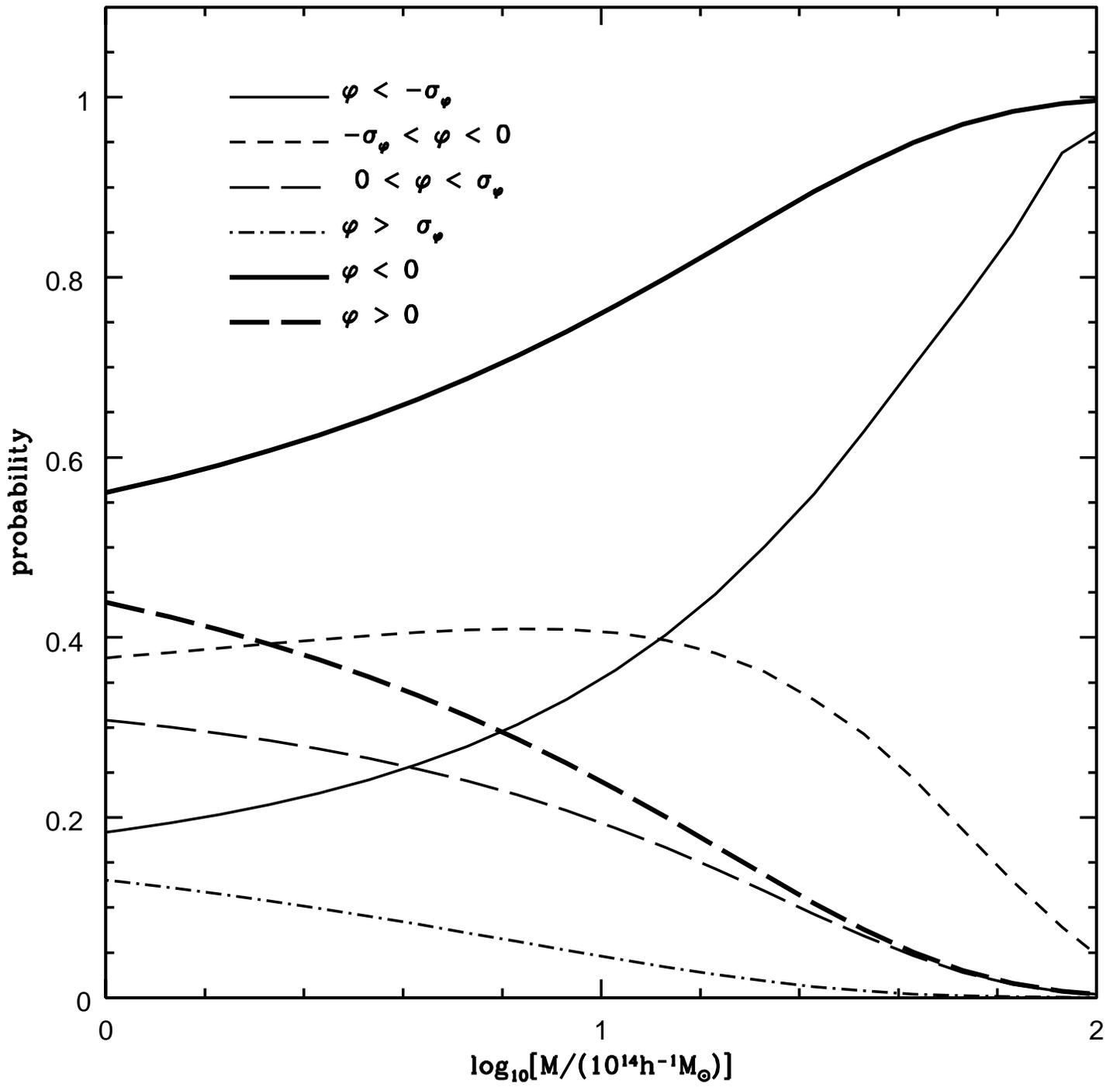
\end{document}